\documentclass[onecolumn,preprintnumbers,aps,superscriptaddress,nofootinbib,prd,notitlepage]{revtex4-1}
\usepackage[colorlinks=true,linkcolor=red,citecolor=blue]{hyperref}
\usepackage{color}
\usepackage{amsmath}
\usepackage{graphicx}
\usepackage{scalefnt}
\usepackage{grffile}
\usepackage{slashed}

\allowdisplaybreaks

\newcommand{\beq}{\begin{eqnarray}}
\newcommand{\eeq}{\end{eqnarray}}
\newcommand{\non}{\nonumber\\ }

\begin{document}
	
\title{Three-loop QCD matching of  the flavor-changing scalar current involving the heavy charm and bottom quark}
\author{Wei Tao~\footnote{taowei@njnu.edu.cn}}
\author{ Ruilin Zhu~\footnote{Corresponding author: rlzhu@njnu.edu.cn} }
\author{Zhen-Jun Xiao~\footnote{Corresponding author: xiaozhenjun@njnu.edu.cn} }
\affiliation{Department of Physics and Institute of Theoretical Physics, Nanjing Normal University, Nanjing, Jiangsu 210023, China}

\date{\today}
\vspace{0.5in}
	
\begin{abstract}
We compute the matching coefficient between the quantum chromodynamics (QCD) and the non-relativistic QCD ( NRQCD)  for the flavor-changing scalar current involving the heavy charm and bottom quark,
up to the three-loop order within the NRQCD factorization.
For the first time, we obtain the analytical expressions for the three-loop  renormalization constant $\tilde{Z}_s(x,R_f)$ and the corresponding anomalous dimension $\tilde{\gamma}_s(x,R_f)$
for	the NRQCD scalar current with the two heavy  bottom and charm quark.  We  present the precise numerical results for those relevant coefficients $(C_{FF}(x_0), \cdots, C_{FBB}(x_0))$
 with an accuracy of about thirty digits.
The three-loop QCD correction turns out to be significantly 	large. The obtained  matching coefficient  $C_s(\mu_f,\mu,m_b,m_c)$ is helpful to analyze the threshold behaviours
when two different heavy quarks are close to each other and form the double heavy $B_c$ mesons.
\end{abstract}
	
\pacs{12.38 Bx, 12.39 St, 13.85. Ni }
\maketitle

\section{Introduction}

Heavy quark system provides a unique window to understand the perturbative and nonperturbative nature of  Quantum Chromodynamics (QCD) theory.
The key quantity of the heavy quark system is the heavy quark mass with $m_Q\gg \Lambda_{QCD}$, which is naturally employed to distinguish the perturbative and nonperturbative
interactions.  For the two heavy quark system such as heavy quarkonium and the threshold production of top quark pair,
the non-relativistic QCD (NRQCD) effective theory is a very successful approach to separate the long-distance nonperturbative interactions and the short-distance perturbative interactions~\cite{Bodwin:1994jh}.
In this effective theory, the QCD observable can be further expanded into the NRQCD effective operator matrix elements with the corresponding matching coefficients.
The matching coefficients can be calculated order by order and can be in series of two small parameters, i.e. the strong coupling constant $\alpha_{s}$ and the quark relative velocity $v$.

Up to now, various  matching coefficients for the two heavy quark currents,  involving the $b\bar{b}$, $\bar{b}c $ and $c\bar{c}$ systems,  have been computed to higher-order accuracy within the NRQCD effective theory.
Taking the double-heavy $B_c^+ =\bar{b}c $ meson as an example, the next-to-leading order (NLO) matching coefficient for the axial-vector current was first obtained in 1995~\cite{Braaten:1995ej}.
Then, the NLO matching coefficient for the vector current was calculated in 1999~\cite{Hwang:1999fc}.
Later on,  the  NLO QCD corrections combined with the higher-order relativistic corrections for   the axial-vector current  and  the vector current are investigated in Ref.~\cite{Lee:2010ts}.
The approximate results of the next-to-next-to-leading order (NNLO) QCD correction for the   axial-vector current matching coefficient was first obtained in 2003~\cite{Onishchenko:2003ui}.
However, the complete analytical expression of the two-loop matching coefficient for the axial-vector current became available in 2015~\cite{Chen:2015csa}.
This year, the next-to-next-to-next-to-leading order (N$^3$LO) QCD correction for the matching coefficient of the axial-vector current  was first numerically calculated in Ref.~\cite{Feng:2022ruy}.
Very recently, the two-loop matching coefficient of the vector current  was evaluated by the authors of this work~\cite{Tao:2022qxa}.
After that, the three-loop matching coefficient of the vector current  was also achieved in Ref.~\cite{Sang:2022tnh}.

For the case of heavy quark currents with equal masses, the matching coefficient has been evaluated within the NRQCD factorization frame at high order by various literature.
For example, the NNLO QCD correction can be found in Ref.~\cite{Kniehl:2006qw}, and the N$^3$LO QCD correction can be found in Refs.~\cite{Piclum:2007an,Egner:2022jot}.
For more higher-order calculations about matching coefficients for doubly heavy quark systems, one can see the following literature ~\cite{Grozin:2007fh,Broadhurst:1994se,Marquard:2006qi,Egner:2021lxd,Beneke:1997jm,Marquard:2009bj,Marquard:2014pea,Kniehl:2002yv,Chen:2017soz,Piclum:2007an,Egner:2022jot,Feng:2022vvk,Tao:2022yur,Zhu:2017lqu,Zhu:2017lwi,Kiselev:1998wb,Lee:2018rgs,Bell:2010mg,Czarnecki:1997vz,Beneke:1997jm}.

However, the matching coefficient for the flavor-changing scalar current involving the heavy bottom and charm quark has not yet been found in previous works.
Considering the state-of-the-art theoretical accuracy of  the matching coefficients for other heavy quark currents, we will compute  the  matching coefficient for the flavor-changing heavy quark scalar current of the $B_c$ meson
up to the three-loop QCD corrections within the NRQCD factorization frame.
The novel calculation is helpful to analyze the threshold behaviours when two different heavy quarks such as the bottom and the charm quark are close to each other.
In addition, the three-loop QCD matching procedure for the heavy flavor-changing  scalar current of the $B_c$ meson provides a window to  check the convergence of the perturbative series of the NRQCD effective theory.

The rest of the paper is arranged as follows.
In Sec.~\ref{Matchingformula}, we introduce the matching formula between the full QCD theory and the NRQCD effective theory.  We  present the analytical results of  the three-loop  scalar current renormalization constant and the corresponding anomalous dimension in the NRQCD effective theory. We also discuss the matching and running equations for the strong coupling constant $\alpha_{s}$.
In Sec.~\ref{Calculationprocedure},  we give our calculation procedure and the projection for the heavy flavor-changing scalar current.
In Sec.~\ref{Results}, we present our numeric results of the matching coefficient up to the three-loop order accuracy. Finally, we summarize  in Sec.~\ref{Summary}.

\section{Matching formula ~\label{Matchingformula}}

The heavy flavor-changing scalar current involving $\bar{b}$ and $c$ quark in the full QCD can be written as $j_s(y)=\bar{\Psi}_b(y) \Psi_c(y)$, which can be further expanded into the NRQCD effective scalar current
\beq
\tilde{j}_s(y)=-\frac{\chi^\dag_b(y) \vec{p}\cdot \vec{\sigma}\psi_c(y)}{2m_{\rm red}},
\eeq
with the reduced heavy quark mass $m_{\rm red}=m_b m_c/(m_b+m_c)$ and the quark relative momentum $p$ at the leading order of quark relative velocity, according
to the definition as given  in Ref.~\cite{Kniehl:2006qw}. $\psi$ is the two-component Pauli spinor field that annihilates a heavy quark, while $\chi$ is the two-component Pauli
spinor field that creates a heavy anitiquark.
The matching coefficient can be determined through the conventional perturbative matching procedure. Namely, one  performs renormalization for  the on-shell vertex functions in both the perturbative QCD and the
NRQCD sides, then solves the matching coefficient  order by order in $\alpha_s$.

The matching formula with renormalization procedure reads
\beq
\label{matchf}
\sqrt{Z_{2,b} Z_{2,c} } \,Z_s  \, \Gamma_s^0 =\mathcal{C}_s(\mu_f,\mu,m_b,m_c) \, \sqrt{\widetilde{Z}_{2,b} \widetilde{Z}_{2,c} } \,
{\widetilde Z}_s^{-1} \, \widetilde{\Gamma}_s^0 + {\mathcal O}(v^2),
\eeq
where  the left hand part of the equation represents the renormalization of the full QCD current while the right hand part represents the renormalization of the NRQCD current.
The term ${\mathcal O}(v^2)$ denotes the higher order relativistic corrections in powers of the heavy quark relative velocity $v$ between the bottom quark  $\bar{b}$ and the charm quark $c$.
 $\Gamma_s^0$ ($\widetilde{\Gamma}_s^0$) denotes the on-shell unrenormalized heavy flavor-changing current vertex function in the QCD (NRQCD) theory, and the leading-order (LO) matching
 coefficient is normalized into 1.
  In this paper, we will consider  higher-order QCD corrections up to ${\mathcal O}(\alpha_s^3)$ but at the lowest order in heavy quark relative velocity $v$~\cite{Marquard:2014pea,Feng:2022vvk}.
 $Z_s$ is the QCD  current renormalization constant in on-shell ($\mathrm{OS}$) scheme, i.e., $Z_s=\frac{m_b Z_{m,b}+m_c Z_{m,c}}{m_b+m_c}$.
 And $\tilde{Z}_s$ is the renormalization constant of the NRQCD effective current  in the modified-minimal-subtraction ($\mathrm{\overline{MS}}$) scheme. $Z_{2}$ and $Z_{m}$ are QCD on-shell quark field and mass renormalization constants, respectively.
 The three-loop analytical results of  the on-shell quark field and mass  renormalization constants  allowing for two different non-zero quark masses  can be found in literature~\cite{Bekavac:2007tk,Marquard:2016dcn,Fael:2020bgs}, which can be  evaluated to high numerical precision  with the package {\texttt{PolyLogTools}}~\cite{Duhr:2019tlz}.
 The QCD    coupling  $\mathrm{\overline{MS}}$   renormalization constant can be found in literature~\cite{Mitov:2006xs,Chetyrkin:1997un,vanRitbergen:1997va}.
 The NRQCD on-shell quark field renormalization constants $\tilde{Z}_{2,b}=\tilde{Z}_{2,c}=1$, since all light particles in NRQCD  are massless. The matching coefficient $\mathcal{C}_s(\mu_f,\mu,m_b,m_c)$ depends on the
 NRQCD factorization scale $\mu_f$ and the QCD  renormalization scale $\mu$ in a finite order QCD correction calculation.

After implementing the quark field, quark mass and  the QCD coupling constant renormalization,  the QCD vertex function  gets rid of ultra-violet(UV) poles, while still contains  uncancelled infra-red(IR) poles starting from order $\alpha_s^2$.  The remaining IR poles in QCD should be exactly  cancelled by the UV divergences of ${\widetilde Z}_s$ in NRQCD, which  renders the matching coefficient finite.  With the aid of the obtained high-precision numerical results, combined with the features of the NRQCD  current renormalization constants investigated in other known literature~\cite{Feng:2022vvk,Feng:2022ruy,Sang:2022tnh,Egner:2022jot},  we have successfully reconstructed the exact analytical expression of the NRQCD  renormalization constant for the flavor-changing heavy quark scalar current through numerical fitting recipes~\cite{ferguson1999analysis,abramowitz1964handbook,Duhr:2019tlz}. Here we directly present the final result  as following
\beq
\widetilde{Z}_s\left(x, {\mu^2_f\over m_b m_c } \right) &=& 1+\left(\frac{\alpha_{s}^{\left(n_{l}\right)}\left(\mu_f\right)}{\pi}\right)^{2} \widetilde{Z}_s^{(2)}(x)
+\left(\frac{\alpha_{s}^{\left(n_{l}\right)}\left(\mu_f\right)}{\pi}\right)^{3} \widetilde{Z}_s^{(3)}\left(x,\frac{\mu^2_f}{m_bm_c} \right)  +\mathcal{O}(\alpha_s^4),  \label{eq:Zs01}
\eeq
where the coefficients $ \widetilde{Z}_s^{(2)}(x)$ and $\widetilde{Z}_s^{(3)}\left(x,\frac{\mu^2_f}{m_bm_c} \right)$ are of the following form
\beq
 \widetilde{Z}_s^{(2)}(x) &=& \pi^{2}C_{F}  \frac{1}{\epsilon}\left(\frac{3x^2+10x+3}{24\left(1+x\right)^2}C_{F}+\frac{1}{24}C_{A}\right), \label{eq:f2epx}
\eeq
\beq
\widetilde{Z}_s^{(3)}\left(x,\frac{\mu^2_f}{m_bm_c} \right) &=& \pi^{2}C_{F} \Bigg\{ \frac{C_F^2}{\epsilon}\left (\frac{57x^2+146x+57}{216(x+1)^2}-\frac{\ln2}{3}\right)  -  \frac{C_F C_A}{\epsilon^2} \frac{11 x^2+41 x+11}{216 (x+1)^2 } \non
&& + \frac{C_FC_A}{1296 \epsilon }\bigg[ \frac{379 x^2+1086 x+379}{(x+1)^2}-72 \ln2-\frac{9 (5 x+11) }{x+1}\ln x \non
&&  +144 \ln (x+1)+\frac{9 \left(11 x^2+28 x+11\right) }{(x+1)^2}\ln \frac{\mu _f^2}{m_b	m_c} \bigg] \non
&& + C_A^2  \Bigg[\frac{-1}{48 \epsilon ^2}+\frac{1}{648 \epsilon}\left(34+72 \ln2-9 \ln x+18 \ln (x+1)+9 \ln \frac{\mu _f^2}{m_b m_c} \right)\Bigg] \non
&& + C_F T_F n_l\bigg[\frac{3 x^2+10	x+3}{108\epsilon ^2 (x+1)^2 }-\frac{21 x^2+74 x+21}{324 \epsilon (x+1)^2 }\bigg] +C_A  T_F n_l \bigg[\frac{1}{108 \epsilon ^2}-\frac{53}{1296 \epsilon }\bigg]\Bigg\} ,  \label{eq:f3epx}
\eeq
where $C_F=4/3,C_A=3,T_F=1/2$, and the parameter  $x =m_c/m_b$ representing the ratio of the heavy charm and bottom quark mass.

The corresponding anomalous dimension $\tilde{\gamma}_{s}$ for the NRQCD scalar current  is related to $\tilde{Z}_s$ by~\cite{Groote:1996xb,Kiselev:1998wb,Henn:2016tyf,Fael:2022miw,Grozin:2015kna,Ozcelik:2021zqt}
\beq
\tilde{\gamma}_s\left(x, {\mu^2_f\over m_b m_c } \right) &\equiv & {d \ln \widetilde{Z}_s \over d \ln \mu_f }  \equiv \frac{-2\, \partial{\tilde{Z}_s^{[1]}}}{\partial\ln\alpha_s^{(n_l)}(\mu_f)} \non
&=& \left(\frac{\alpha_s^{(n_l)} 	\left(\mu_f\right)}{\pi}\right)^2 \tilde{\gamma}_s^{(2)}(x ) +\left(\frac{\alpha_s^{\left(n_l\right)	}
\left(\mu_f\right)}{\pi}\right)^3 \tilde{\gamma}_s^{(3)}\left(x, {\mu^2_f\over m_b m_c }\right)+\mathcal{O}(\alpha^4_s), \label{eq:Gammas}
\eeq
where $\tilde{Z}_{s}^{[1]}$ denotes the coefficient of the $\frac{1}{\epsilon}$ pole in $\tilde{Z}_s$, and the  NRQCD factorization scale $\mu_f$ is used because both  $\tilde{Z}_{s}$ and  $\tilde{\gamma}_{s}$ are defined
in the NRQCD effective theory. The explicit expressions of the coefficients $\tilde{\gamma}_s^{(2)}(x )$ and $\tilde{\gamma}_s^{(3)}\left(x, {\mu^2_f\over m_b m_c }\right)$ in Eq.~(\ref{eq:Gammas}) are of the form of
\beq
\tilde{\gamma}_s^{(2)}(x )&=& -\pi^2C_F\left[  C_F \frac{3 x^2+10	x+3}{6(x+1)^2} +\frac{C_A}{6} \right], \label{eq:Gammas2} \\
\tilde{\gamma}_s^{(3)}\left(x, {\mu^2_f\over m_b m_c }\right)&=&  \pi^2C_F\Bigg\{  C_F^2 \left[-\frac{57 x^2+146 x+57}{36 (x+1)^2}+2 \ln 2 \right] \non
&&\hspace{-2cm}+ C_A^2 \left[-\frac{17}{54}-\frac{2 }{3}\ln 2+\frac{1}{12}\ln x-\frac{1}{6} \log (x+1)-\frac{1}{12} \ln \frac{\mu_f^2}{m_b m_c}\right] \non
&&\hspace{-2cm}+ C_F C_A \bigg[-\frac{379 x^2+1086 x+379}{216 (x+1)^2}+\frac{\ln 2 -2\ln(x+1)}{3} +\frac{5 x+11 }{24 (x+1)}\ln x-\frac{11 x^2+28 x+11}{24 (x+1)^2}\ln\frac{\mu_f^2}{m_b m_c}\bigg] \non
&&\hspace{-2cm}+ C_F T_F n_l \frac{21 x^2+74 x+21}{54 (x+1)^2}+\frac{53}{216} C_A T_F n_l \Bigg\}. \label{eq:Gammas3}
\eeq
From above expressions, one can see that both of $\tilde{Z}_{s}$ and  $\tilde{\gamma}_{s}$ explicitly depend on the NRQCD factorization scale $\mu_f$, which is a feature found in other NRQCD currents~\cite{Marquard:2014pea,Egner:2022jot,Feng:2022vvk,Feng:2022ruy,Sang:2022tnh}.
Note that the above three-loop expressions of $\widetilde{Z}_{s}$ and $\tilde{\gamma}_{s}$ for  the flavor-changing scalar current involving the heavy charm and bottom quark are known for the first time.
The obtained $\widetilde{Z}_{s}$ and $\tilde{\gamma}_{s}$ have been checked with several different values of $m_b$ and $m_c$.
To verify the correction of our results, on the one hand, one can check that  above $\widetilde{Z}_s\left(\tilde{\gamma}_{s}\right)$  is  symmetric under the combined exchange of $m_b\leftrightarrow m_c$ and
$n_b\leftrightarrow n_c$.  On the other hand, in the equal quark mass case of $x=1$, our $\widetilde{Z}_s$ and $\tilde{\gamma}_{s}$ are in full agreement with the known results
as given in Refs.~\cite{Kniehl:2006qw,Piclum:2007an,Egner:2022jot}.

In our calculation, we include the contributions from the loops of charm quark and bottom quark in the full QCD, which however are decoupled in the NRQCD.
To match the QCD with the NRQCD, one need apply the decoupling relation~\cite{Chetyrkin:2005ia,Bernreuther:1981sg,Barnreuther:2013qvf,Grozin:2007fh,Ozcelik:2021zqt} of $\alpha_s$, i.e., the coupling constants
$\alpha_s^{(n_l+1)}(\mu)$ in QCD (with $n_l+1$ flavours) and $\alpha_s^{(n_l)}(\mu)$ in the NRQCD (with $n_l$ light flavours) are related by~\cite{Grozin:2007fh}
{\small
\beq
\frac{\alpha_s^{(n_l+1)}(\mu)}{\pi} &=& \frac{\alpha_s^{(n_l)}(\mu)}{\pi} + \left( \frac{\alpha_s^{(n_l)}(\mu)}{\pi} \right)^2 T_F \bigg[ \frac{1}{3} L + \left( \frac{1}{6} L^2 + \frac{1}{36} \pi^2 \right) \epsilon
+ \left( \frac{1}{18} L^3 + \frac{1}{36} \pi^2 L - \frac{1}{9} \zeta_3 \right) \epsilon^2 + \mathcal{O}{(\epsilon^3)} \bigg]\non
&+& \left( \frac{\alpha_s^{(n_l)}(\mu)}{\pi} \right)^3 T_F \bigg\{ \left( \frac{1}{4} L +\frac{15}{16} \right) C_F + \left( \frac{5}{12} L - \frac{2}{9} \right) C_A
+ \frac{1}{9} T_F L^2 + \bigg[ \left( \frac{1}{4} L^2 + \frac{15}{8} L + \frac{1}{48} \pi^2 + \frac{31}{32} \right) C_F \non
&+&  \left( \frac{5}{12} L^2 - \frac{4}{9} L + \frac{5}{144} \pi^2 + \frac{43}{108} \right) C_A + \left( \frac{1}{9} L^3 + \frac{1}{54} \pi^2 L \right) T_F \bigg] \epsilon
+ \mathcal{O}{(\epsilon^2)} \bigg\} + \mathcal{O}{(\alpha_s^4)}\,, \label{eq:alphasnlp1}
\eeq
where $L =\ln(\mu^2/m_Q^2)$ and $m_Q$ is the on-shell mass of the decoupled heavy quark.

Besides, we  can evolve the strong coupling from the scale $\mu_f$ to the scale $\mu$ with  renormalization group running equation~\cite{Abreu:2022cco} in $D=4-2 \epsilon$ dimensions  as following
\begin{align}\label{asrun1}
\alpha_s^{(n_l)}\left(\mu_f\right)=
\alpha_s^{(n_l)}\left(\mu\right)\left(\frac{\mu}{\mu_f}\right)^{2\epsilon}
\bigg[1+\frac{\alpha_s^{(n_l)}\left(\mu\right)}{\pi}\frac{\beta_0^{(n_l)}}{4\epsilon}\left(\left(\frac{\mu}{\mu_f}\right)^{2\epsilon}-1\right)\bigg] + \mathcal{O}{(\alpha_s^3)}.
\end{align}
To calculate the values of the strong coupling,  we  also use the renormalization group running equation~\cite{Chetyrkin:2000yt} in $D=4$ dimensions as
\begin{align}\label{asrun2}
\frac{\alpha_s^{(n_l)}\left(\mu\right)}{4\pi}=
\frac{1}{\beta_0^{(n_l)}L_{\Lambda}}-\frac{b_1 \ln L_{\Lambda}}{\left(\beta_0^{(n_l)}L_{\Lambda}\right)^2}+\frac{b_1^2(\ln^2 L_{\Lambda}-\ln L_{\Lambda}-1)+b_2}{\left(\beta_0^{(n_l)}L_{\Lambda}\right)^3}+\mathcal{O}{\left(\left(\frac{1}{L_{\Lambda}}\right)^4\right)},
\end{align}
where $L_{\Lambda}=\ln\left(\mu^2/{\Lambda_{QCD}^{(n_l)}}^2\right)$ and $b_i=\beta_i^{(n_l)}/{\beta_0^{(n_l)}}$.
At the  one-, two- and three-loop level, the coefficients of the QCD $\beta$ function  are of the form of
\beq
 \beta_0^{(n_l)} &=& \frac{11}{3} C_A- \frac{4}{3} T_F n_l, \non
 \beta_1^{(n_l)} &=& \frac{34}{3} C_A^2- \frac{20}{3} C_A T_F n_l - 4 C_F T_F n_l,  \non
 \beta_2^{(n_l)} &=& \frac{2857}{54} C_A^3- \left ( \frac{1415}{27} C_A^2+\frac{205}{9} C_A C_F-2C_F^2\right ) T_F n_l+ \left ( \frac{158}{27} C_A+\frac{44}{9}  C_F \right ) T_F^2 n_l^2.
\eeq
In our numerical evaluations,  $n_b=n_c=1$ and $n_l=3$ are fixed through the decoupling region from $\mu=1\,\mathrm{GeV}$ to $\mu=6.25\,\mathrm{GeV}$,
and the typical QCD scale $\Lambda_{QCD}^{(n_l=3)}=0.3344\mathrm{GeV}$  is  determined using the three-loop formula with the aid of the package {\texttt{RunDec}}~\cite{Chetyrkin:2000yt,Schmidt:2012az,Deur:2016tte,Herren:2017osy}  by inputting the initial value $\alpha_s^{(n_f=5)}\left(m_Z=91.1876\mathrm{GeV}\right)=0.1179$.

\section{Calculation procedure~\label{Calculationprocedure}}

Our high-order calculation  consists of the following steps.
First, we use {\texttt{FeynCalc}}~\cite{Shtabovenko:2020gxv}  to obtain Feynman diagrams and corresponding Feynman amplitudes.
By {\texttt{\$Apart}}~\cite{Feng:2012iq}, we decompose  every Feyman amplitude  into several Feynman integral families.
Second, we use {\texttt{FIRE}}~\cite{Smirnov:2019qkx}/{\texttt{Kira}}~\cite{Klappert:2020nbg}/{\texttt{ FiniteFlow}}~\cite{Peraro:2019svx} based on Integration by Parts (IBP)~\cite{Chetyrkin:1981qh}
to reduce every Feynman integral family to master integral family.
Third, based on symmetry among different integral families and using  {\texttt{Kira}}+{\texttt{FIRE}}+{\texttt{Mathematica\,code}}, we can realize  integral reduction among different integral families,
and further on, the reduction from all of master integral families to the minimal set~\cite{Fael:2020njb} of master  integral families.
Last, we use {\texttt{AMFlow}}~\cite{Liu:2022chg}, which is a proof-of-concept implementation of the auxiliary mass flow method~\cite{Liu:2017jxz}, equipped with {\texttt{Kira}}~\cite{Klappert:2020nbg}/{\texttt{FiniteFlow}}~\cite{Peraro:2019svx} to calculate the minimal set of master integral families.

In order to obtain the finite results of  the high-order QCD corrections, one has to perform the  conventional renormalization procedure~\cite{Chen:2015csa,Kniehl:2006qw,Bonciani:2008wf,Davydychev:1997vh}.
Equivalently, we can also use diagrammatic renormalization method~\cite{deOliveira:2022eeq}  with the aid of the package {\texttt{FeynCalc}}~\cite{Shtabovenko:2020gxv},
which at N$^3$LO sums contributions from three-loop diagrams and four  kinds of counter-term diagrams, i.e.,  tree diagram   inserted with one $\alpha_s^3$-order counter-term vertex,
one-loop diagram inserted with one $\alpha_s^2$-order counter-term vertex, one-loop diagram inserted with two $\alpha_s$-order counter-term vertexes,
two-loop diagrams inserted with one $\alpha_s$-order counter-term vertex. Our final finite results by these two renormalization methods are in agreement with each other.

We want to mention that all contributions up to NNLO have been evaluated for general gauge parameter $\xi$ and the NNLO results for the scalar current matching coefficient are all independent of $\xi$,
which constitutes an important check on our calculation.
At N$^3$LO, we work in Feynman gauge. By FeynCalc, there are 1, 1, 13, 268 bare Feynman diagrams for the QCD vertex function with the flavor-changing heavy quark scalar current at tree, one-loop, two-loop, three-loop orders in $\alpha_s$, respectively.
Some representative Feynman diagrams up to  three loops  are displayed in Fig.~\ref{fig:fig1} and Fig.~\ref{fig:fig2}.
In the calculation of multi-loop diagrams, we have allowed for $n_b$ bottom quarks with mass $m_b$, $n_c$ charm quarks  with mass $m_c$ and $n_l$ massless quarks appearing in the quark loop.
Physically,  $n_b=n_c=1$ and $n_l=3$.
To facilitate our calculation, we take full advantage of computing numerically. Namely, before generating amplitudes, $m_b$ and $m_c$ are chosen
to be particular rational number values\cite{Bronnum-Hansen:2021olh,Chen:2022vzo,Chen:2022mre}.
Following the literature~\cite{Kniehl:2006qw}, we employ the projector constructed for the flavor-changing heavy quark scalar current to obtain intended QCD amplitudes,
which means one need extend the scalar current projector with equal heavy quark masses in Eq.~(8) of Ref.~\cite{Kniehl:2006qw} to the different heavy quark masses case.
Adopting the same notation of Ref.~\cite{Kniehl:2006qw}, we choose $q_1=\frac{m_c}{m_b+m_c}q+p$ and $q_2=\frac{m_b}{m_b+m_c}q-p$ denoting the on-shell charm and bottom momentum, respectively,
and present the projector for the flavor-changing heavy quark scalar current as
\beq
P_{(s)} &=& \frac{1}{2(m_b+m_c)^2} \Bigg\{ \frac{m_c}{m_b+m_c}\left(\frac{m_c}{m_b+m_c}\slashed{q} + m_c \right) {\bf 1} \left(\frac{m_b}{m_b+m_c}\slashed{q} + m_b\right) \non
          && +\frac{m_b}{m_b+m_c}\left(-\frac{m_c}{m_b+m_c}\slashed{q} + m_c \right) {\bf 1} \left(-\frac{m_b}{m_b+m_c}\slashed{q} + m_b\right) \non
          &&  +\frac{2 m_b m_c}{m_b+m_c} \left(\frac{m_c}{m_b+m_c}\slashed{q} + m_c\right) \frac{\slashed{p}}{p^2} \left(-\frac{m_b}{m_b+m_c}\slashed{q} + m_b\right) \Bigg\},
\eeq
where the small momentum $p$ refers to  relative movement  between the bottom and charm, $q$ represents the total momentum of the bottom and charm, $q_1^2=m_c^2$, $q_2^2=m_b^2$, $q^2=(m_b+m_c)^2+\mathcal{O}{(p^2)}$, $q\cdot p=0$.

To match with the NRQCD, one need extract the contribution from the hard region in the full QCD amplitudes for the scalar current, which means one need first introduce the small relative momentum $p$ to momenta in the amplitudes as above and then series expand propagator denominators with respect to $p$ up to $\mathcal{O}{(p)}$ in  the hard region of loop momenta~\cite{Kniehl:2006qw}.
As a result, the number and powers of propagators in Feynman integrals constituting the amplitudes for the scalar current will remarkably increase compared with  the vector current case,
the zeroth component of the axial-vector current and the pseudoscalar current case.
In our practice, the total number  of propagators in a three-loop Feynman integral family is 12.
In our calculation, the most difficult thing is the reduction from Feynman integrals    with rank 5, dot 4, and 12  propagators  to the master integrals.
By trial and  error, we find it is more  appropriate for  Fire6~\cite{Smirnov:2019qkx} to deal with this problem than {\texttt{Kira}}~\cite{Klappert:2020nbg} or {\texttt{FiniteFlow}}~\cite{Peraro:2019svx}.
After using  {\texttt{Kira}}+{\texttt{FIRE}}+{\texttt{Mathematica\,code}} to achieve the minimal set of the master integral families based on symmetry among different integral families,
the number of three-loop master integral families is reduced from 829  to 26, meanwhile the number of three-loop master integrals is reduced from 13251  to 300.

\begin{figure}[thb]
\center{\includegraphics*[scale=0.6]{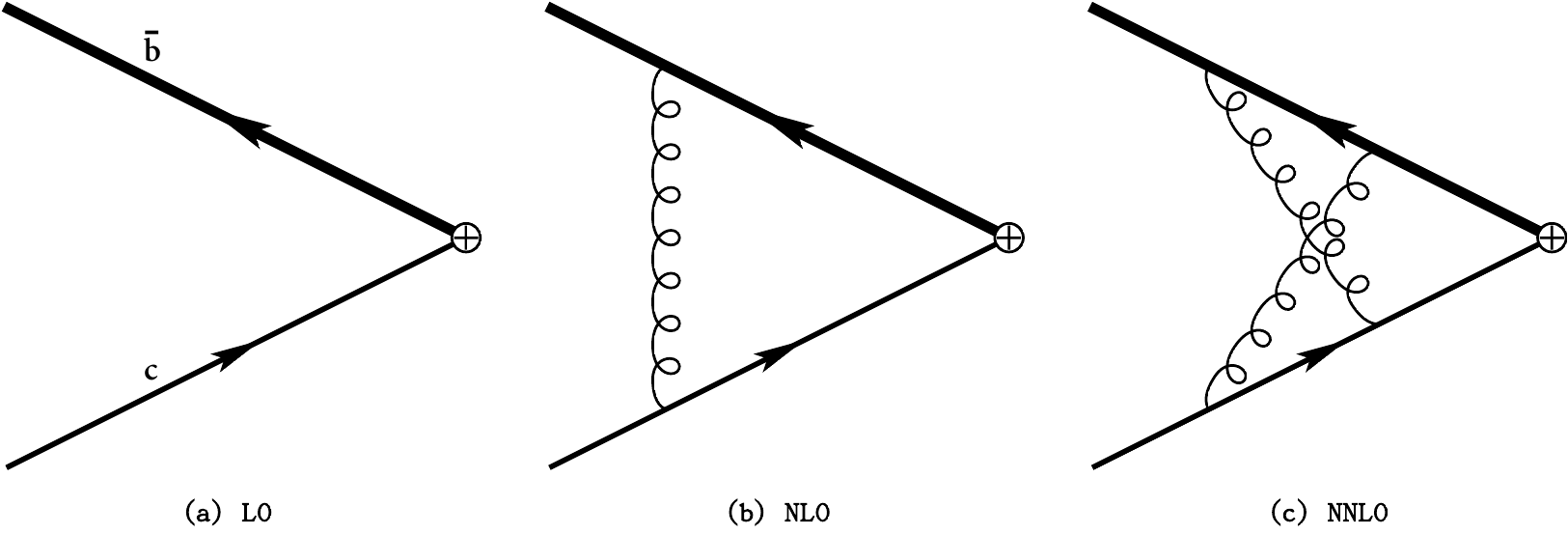}
\caption {\label{fig:fig1} Typical Feynman diagrams 	for the QCD vertex function with the $\bar{b}c $ system
up to two-loop order.	 The cross ``$\bigoplus$'' implies the insertion of the flavor-changing heavy quark scalar current.	} }
\end{figure}

\begin{figure}[thb]
\center{\includegraphics*[scale=0.9]{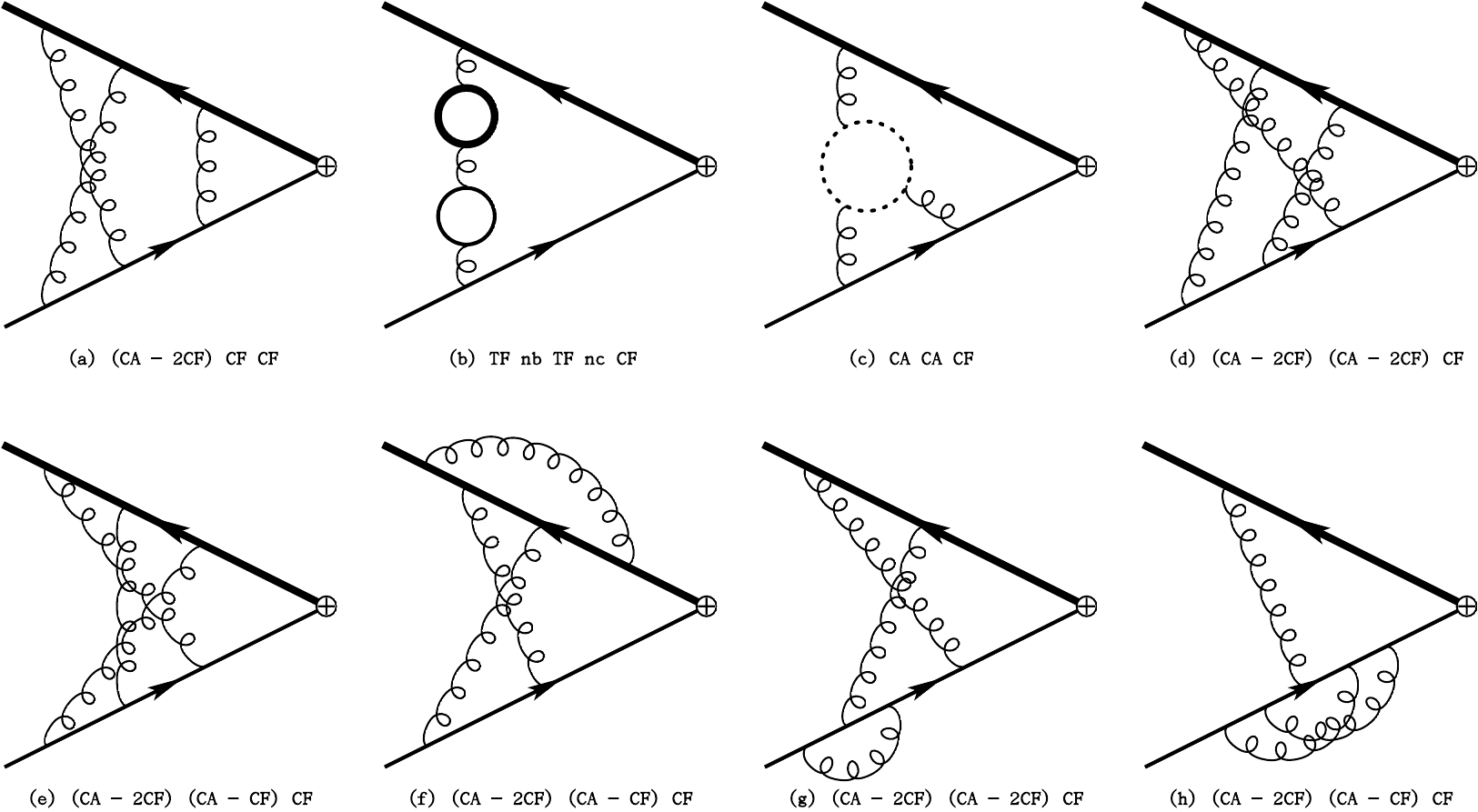}
\caption {\label{fig:fig2} Typical  Feynman diagrams  labelled with the corresponding color factor for the QCD vertex function with the flavor-changing heavy quark scalar current at three-loop order.	
The cross ``$\bigoplus$'' implies the insertion of the scalar current. The thickest solid	closed circle represents the bottom quark loop, and the other solid	closed circle represents
the charm quark loop. The dotted closed circle represents the ghost loop.}}
\end{figure}

\section{Results~\label{Results}}

Following Refs.~\cite{Feng:2022vvk,Feng:2022ruy,Sang:2022tnh}, the dimensionless matching coefficient $\mathcal{C}_s$ for the flavor-changing scalar current involving the heavy bottom and charm quark
 can be decomposed as:
{\small
\beq
\mathcal{C}_s(\mu_f,\mu,m_b,m_c) &=&1+\frac{\alpha_s^{\left(n_l\right)}\left(\mu\right)}{\pi} \mathcal{C}^{(1)}(x)+\left(\frac{\alpha_s^{\left(n_l\right)}\left(\mu\right)}{\pi}\right)^2
\left(\mathcal{C}^{(1)}(x)\frac{\beta_0^{(n_l)}}{4}\text{ln}\frac{\mu^2}{m_b m_c}+\frac{\tilde{\gamma}_s^{(2)}(x)}{2}\ln \frac{\mu_f^2}{m_b m_c}+\mathcal{C}^{(2)}(x)\right) \non
&& \hspace{-1cm}+\left(\frac{\alpha_s^{\left(n_l\right)}\left(\mu\right)}{\pi}\right)^3\Bigg \{  \left(\frac{\mathcal{C}^{(1)}(x)}{16}\beta_1^{(n_l)}+\frac{\mathcal{C}^{(2)}(x)}{2}\beta_0^{(n_l)}\right)\text{ln}\frac{\mu^2}{m_b m_c}
+\frac{\mathcal{C}^{(1)}(x)}{16}{\beta_0^{(n_l)}}^2\ln^2 \frac{\mu^2}{m_b m_c}  \non
&& \hspace{-1cm} -\frac{1}{8}\left(\frac{d\tilde{\gamma}_s^{(3)}\left(x,\frac{\mu_f^2}{m_b m_c}\right)}{d \text{ln}\mu_f}+\beta_0^{(n_l)}\tilde{\gamma}_s^{(2)}(x)\right)\ln^2\frac{\mu_f^2}{m_b m_c}  +\frac{1}{2}\left(\mathcal{C}^{(1)}(x) \tilde{\gamma}_s^{(2)}(x)+\tilde{\gamma}_s^{(3)}\left(x,\frac{\mu_f^2}{m_b m_c}\right)\right)\ln\frac{\mu_f^2}{m_b m_c}  \non
&&  \hspace{-1cm}+\frac{\beta_{0}^{(n_l)}}{4}\tilde{\gamma}_s^{(2)}(x)\ln\frac{\mu_f^2}{m_b m_c}\,\text{ln}\frac{\mu^2}{m_b m_c} + \mathcal{C}^{(3)}(x) \Bigg\} +\mathcal{O}\left(\alpha_s^4\right), \label{eq:Cs01}
\eeq }
where the coefficients $\tilde{\gamma}_s^{(2)}(x)$ and $ \tilde{\gamma}_s^{(3)}\left(x,\frac{\mu_f^2}{m_b m_c}\right)$ have been defined in Eqs.~(\ref{eq:Gammas},\ref{eq:Gammas2},\ref{eq:Gammas3}).
The parameters $\mathcal{C}^{(i)}(x)(i=1,2,3)$ in above equation are  independent of $\ln \mu$ and $\ln \mu_f$ and are the nontrivial parts of $\mathcal{C}_s$ at $\mathcal{O}\left(\alpha_s^i\right)$.
It's also well known that the coefficients $\mathcal{C}_s$ and $\mathcal{C}^{(i)}(x)$ satisfy the following symmetric replacements ~\cite{Braaten:1995ej,Hwang:1999fc,Lee:2010ts,Onishchenko:2003ui,Chen:2015csa,Feng:2022ruy,Sang:2022tnh}:
\beq
\mathcal{C}_s(\mu_f,\mu,m_b,m_c)&=&\mathcal{C}_s(\mu_f,\mu,m_c,m_b)|_{n_b\leftrightarrow n_c}, \label{eq:Cs02} \\
\mathcal{C}^{(i)}(x)&=&\mathcal{C}^{(i)}\left(\frac{1}{x}\right)|_{n_b\leftrightarrow n_c}. \label{eq:Cs03}
\eeq

The one-loop QCD correction to $\mathcal{C}_s$, denoted by $\mathcal{C}^{(1)}(x)$, can be analytically achieved as:
\beq
 \mathcal{C}^{(1)}(x)=\frac{3}{4} C_F \left(\frac{x-1}{x+1}\,\ln x-\frac{2}{3}\right).
\eeq

The  two-loop and three-loop matching coefficients in \eqref{eq:Cs01} are  $\mathcal{C}^{(2)}(x)$ and $\mathcal{C}^{(3)}(x)$, which can be decomposed  in terms of the different color/flavor structures
by following the conventions as being used in Refs. \cite{Marquard:2014pea,Beneke:2014qea,Egner:2022jot,Feng:2022vvk,Feng:2022ruy,Sang:2022tnh}:
\beq
\mathcal{C}^{(2)}(x) &=& C_F^2 \mathcal{C}_{FF}(x)+C_F C_A \mathcal{C}_{FA}(x)+C_F T_F n_b \mathcal{C}_{FB}(x) +C_F T_F n_c \mathcal{C}_{FC}(x) +C_F T_F n_l \mathcal{C}_{FL}(x), \\
\mathcal{C}^{(3)}(x) &=& C^3_F \, \mathcal{C}_{FFF}(x)+C^2_F \,C_A \, \mathcal{C}_{FFA}(x) + C_A^2 C_F \, \mathcal{C}_{FAA}(x)  +  C^2_F T_F n_l\,\mathcal{C}_{FFL}(x) +C_A C_F T_F n_l\,\mathcal{C}_{FAL}(x) \non
&&  + C_F T^2_F n_l\, n_c \, \mathcal{C}_{FCL}(x) + C_FT^2_F \, n_l n_b \, \mathcal{C}_{FBL}(x)  +C_F T^2_F \, n^2_l \, \mathcal{C}_{FLL}(x) + C_F T_F^2 \, n_b \, n_c \, \mathcal{C}_{FBC}(x)  \non
&&  +     C^2_F T_F n_c\,\, \mathcal{C}_{FFC}(x)    + C_A C_F T_F \,n_c\, \mathcal{C}_{FAC}(x)+  C_FT^2_F \, n_c^2 \, \mathcal{C}_{FCC}(x) \non
&&  +   C^2_F T_F\,n_b\, \mathcal{C}_{FFB}(x)+ C_AC_F T_F\,n_b\,\mathcal{C}_{FAB}(x)+C_FT^2_F \, n_b^2 \, \mathcal{C}_{FBB}(x).
\eeq
Due to limited computing resources,
we choose to calculate the matching coefficient $\mathcal{C}_s$ at three rational numerical points: the physical point  $\{m_b=\frac{475}{100}\,\mathrm{GeV}, m_c=\frac{150}{100}\,\mathrm{GeV}\}\left(i.e.,x=x_0=\frac{150}{475}\right)$,   the check  point $\{m_b=\frac{475}{100}\,\mathrm{GeV},m_c=\frac{475}{100}\times\frac{475}{150}\,\mathrm{GeV}\}\left(i.e.,x=\frac{475}{150}\right)$ and the equal mass point $\{m_b=\frac{475}{100}\,\mathrm{GeV},m_c=\frac{475}{100}\,\mathrm{GeV}\}\left(i.e.,x=1\right)$, respectively.
The results $\mathcal{C}_s$ obtained at the physical point and the check  point verify the symmetric features of $\mathcal{C}^{(i)}(x)$ in Eq.~\eqref{eq:Cs03}. Our results $\mathcal{C}_s$ obtained at the equal mass point $x=1$ are consistent with the known  matching coefficient results $\mathcal{C}_s$ for the scalar current with the equal quark mass case in the previous literatures~\cite{Egner:2022jot,Kniehl:2006qw,Piclum:2007an}.
To confirm our calculation,  we have also applied the same calculation procedure to the evaluation of the three-loop matching coefficients $\mathcal{C}_v$, $\mathcal{C}_p$, $\mathcal{C}_{(a,0)}$  for  the flavor-changing heavy quark vector current, the flavor-changing heavy quark  pseudoscalar current,  the zeroth component of the axial-vector current with the flavor-changing heavy quarks, respectively, where our results verify  $\mathcal{C}_p\equiv\mathcal{C}_{(a,0)}$ and our results $\mathcal{C}_v$ and  $\mathcal{C}_{(a,0)}$  agree with the known results in previous literature~\cite{Kniehl:2006qw,Piclum:2007an,Marquard:2014pea,Tao:2022qxa,Egner:2022jot,Feng:2022ruy,Sang:2022tnh}.

In the following, we will present the highly accurate numerical results of $\mathcal{C}^{(2)}(x)$ and $\mathcal{C}^{(3)}(x)$ at the physical heavy quark mass ratio  $x=x_0=1.5/4.75$  with about 30-digit precision.
The various components of ${\cal C}^{(2)}(x_0)$   read:
\beq
\mathcal{C}_{FF}(x_0) &=& -6.96020737354849312657205418357, \non
\mathcal{C}_{FA}(x_0) &=& -4.12970820397051570036738297443,\non
\mathcal{C}_{FB}(x_0) &=& 0.048170796075386686136602545271, \non
\mathcal{C}_{FC}(x_0) &=& 0.268781876466689639100436656736, \non
\mathcal{C}_{FL}(x_0) &=& -0.363661393326874053432216153222.  \label{eq:CFFx0}
\eeq
And the various components of ${\cal C}^{(3)}(x_0)$ read:
\beq
  \mathcal{C}_{FFF}(x_0) &=& -12.6512824902497489841790999287, \non
  \mathcal{C}_{FFA}(x_0) &=& -91.3076763843495687930187876995, \non
  \mathcal{C}_{FAA}(x_0) &=& -67.2034246352357623358462321068,\non
  \mathcal{C}_{FFL}(x_0) &=& 31.12323218543900065296825277243, \non
  \mathcal{C}_{FAL}(x_0) &=& 19.49987491622541782889333507621, \non
  \mathcal{C}_{FCL}(x_0) &=& -0.262383795777090819586462489511, \non
  \mathcal{C}_{FBL}(x_0) &=& -0.0100054359359386271416783335057, \non
  \mathcal{C}_{FLL}(x_0) &=& 0.3393017746199103339986949104357, \non
  \mathcal{C}_{FBC}(x_0) &=& 0.0551829420149711792507691937024, \non
  \mathcal{C}_{FFC}(x_0) &=& 4.4105666464862568415402096694718, \non
  \mathcal{C}_{FAC}(x_0) &=& -0.6861454400278606762848799078670,\non
  \mathcal{C}_{FCC}(x_0) &=& 0.09536045103106728669303211267177,\non
  \mathcal{C}_{FFB}(x_0) &=& 1.1373781611175929139065523549900,\non
  \mathcal{C}_{FAB}(x_0) &=& -0.318775996588438248091557753877,\non
  \mathcal{C}_{FBB}(x_0) &=&  0.00994219408487397025982150577842. \label{eq:CFFFx0}
\eeq
From the numerical results  in Eqs.~(\ref{eq:CFFx0},\ref{eq:CFFFx0}), we find the dominant contributions in $\mathcal{C}^{(2)}(x_0)$ and $\mathcal{C}^{(3)}(x_0)$ come from the components
corresponding to the color structures $C_F^2$, $C_FC_A$, $C_F^2C_A$ and $C_FC_A^2$, and the contributions from the bottom and charm quark loops are negligible.

Fixing the renormalization scale $\mu=m_b=4.75\mathrm{GeV}$, $m_c=1.5\mathrm{GeV}$, and setting the factorization scale $\mu_f=1\,\mathrm{GeV}$, Eq.~(\ref{eq:Cs01}) then reduces to
\beq
\mathcal{C}_s(x_0) & =& 1-0.06727332\left(\frac{\alpha_s^{\left(n_l=3\right)}(m_b)}{\pi}\right)-12.41489\left(\frac{\alpha_s^{\left(n_l=3\right)}(m_b)}{\pi}\right)^2\non
&& -930.1229\left(\frac{\alpha_s^{\left(n_l=3\right)}(m_b)}{\pi}\right)^3+\mathcal{O}(\alpha_s^4). \label{Csasnum}
\eeq
With the values of $\alpha_s^{\left(n_l=3\right)}(\mu)$ calculated by the renormalization group running equation in Eq.~\eqref{asrun2},
we investigate the renormalization scale dependence of the matching coefficient $\mathcal{C}_s$ for scalar current at LO,  NLO,  NNLO and N$^3$LO accuracy in Fig.~\ref{fig:fig3}.

\begin{figure}[thb]
\center{\includegraphics[width=0.6\textwidth]{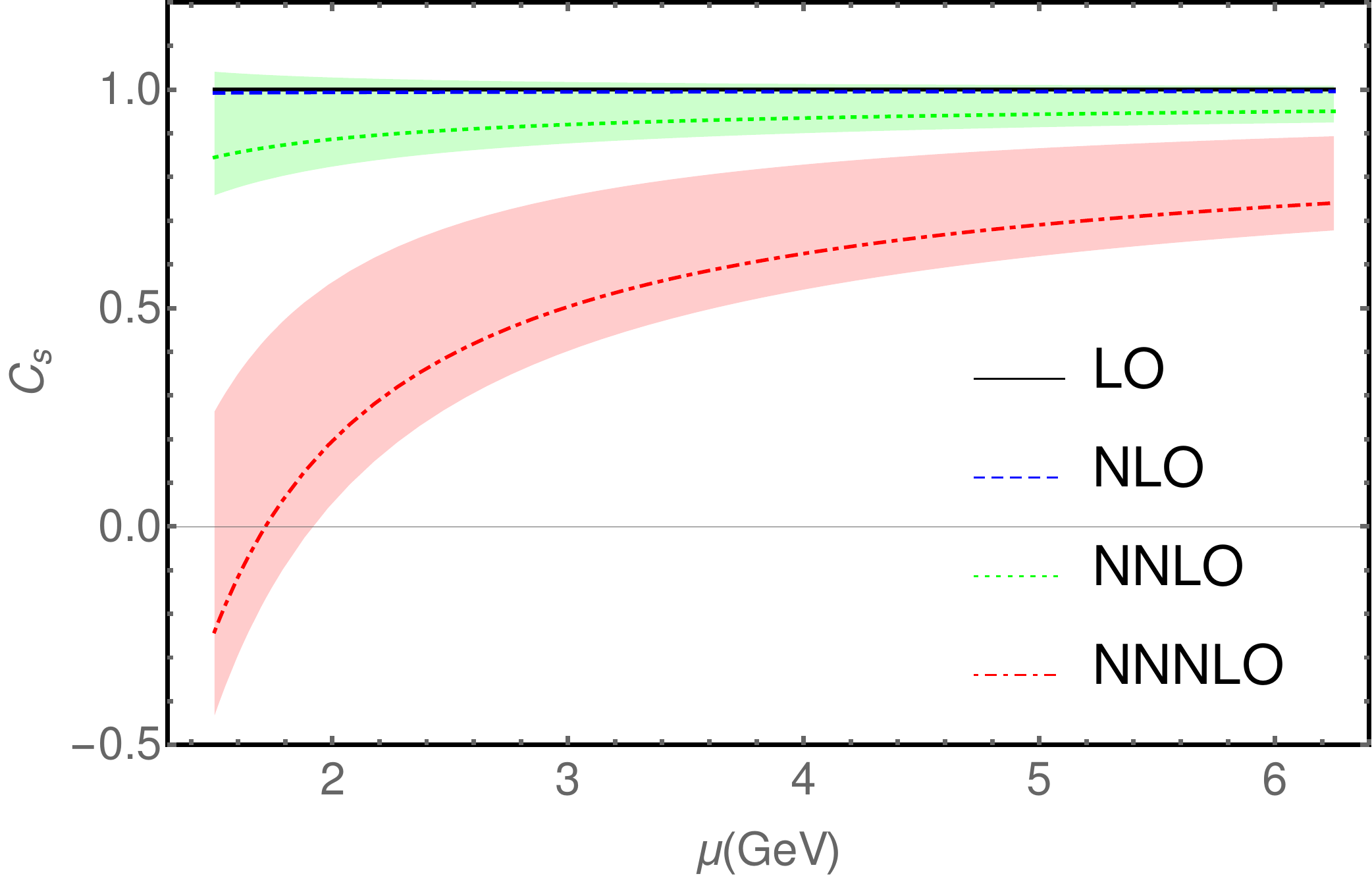}
\caption{\label{fig:fig3} The renormalization scale dependence of the matching coefficient $\mathcal{C}_s$  at the LO,  NLO,  NNLO and N$^3$LO accuracy.
The central values of  the matching coefficient $\mathcal{C}_s$ are calculated inputting the  physical values with $\mu_f=1~\,\mathrm{GeV}$,  $m_b=4.75\mathrm{GeV}$ and $m_c=1.5\mathrm{GeV}$.
The error bands come from varying  $\mu_f$  from   1.5 to 0.4 $\mathrm{GeV}$. }  }
\end{figure}

We also present our precise numerical results of the matching coefficient $\mathcal{C}_s$ for scalar current at LO,  NLO ,  NNLO and N$^3$LO in Tab.~\ref{tab:Csnum}.
The central values of  the matching coefficient $\mathcal{C}_s$ are calculated by inputting the  physical values with $\mu_f=1~\mathrm{GeV}$, $\mu=4.75\mathrm{GeV}$, $m_b=4.75\mathrm{GeV}$ and $m_c=1.5\mathrm{GeV}$.
The errors are estimated by varying $\mu_f$  from  1.5 to 0.4 $\mathrm{GeV}$, $\mu$  from   6.25 to 3 $\mathrm{GeV}$, respectively.

\begin{table}[thb]
\caption{The values of the matching coefficient $\mathcal{C}_s$ for the flavor-changing heavy quark  scalar  current up to N$^3$LO. For details, see the text.  \label{tab:Csnum}}
\renewcommand\arraystretch{2} 		\tabcolsep=0.2cm
		\begin{tabular}{ c c c c c}\hline\hline
			& LO         &  NLO                   & NNLO     & N$^3$LO			\\\hline
			$\mathcal{C}_s$	 & $1$ & $0.99557^{-0+0.00037}_{+0-0.00082}$   &    $0.94186^{-0.03068+0.00873}_{+0.06933-0.02179}$  &     $0.67714^{-0.07379+0.06332}_{+0.18133-0.17414}$			\\		\hline \hline
\end{tabular}\end{table}

From Fig.~\ref{fig:fig3} and Tab. ~\ref{tab:Csnum}, one can find that the higher order QCD corrections have large influence compared with
the NLO correction. Especially, the ${\cal O}(\alpha_s^3)$ correction looks quite sizable, which confirms the breakdown of the convergence at ${\cal O}(\alpha_s^3)$ for other heavy quark currents in previous literatures~\cite{Egner:2022jot,Feng:2022ruy,Sang:2022tnh}.
Note that, at each truncated perturbative order, the matching coefficient $\mathcal{C}_s$  is renormalization-group invariant~\cite{Feng:2022ruy,Sang:2022tnh}, e.g., at N$^3$LO, $\mathcal{C}_s$ obeys the following renormalization-group running invariance:
\begin{align}
&\mathcal{C}_s^{\rm N^3LO}(\mu_f,\mu,m_b,m_c) =\mathcal{C}_s^{\rm N^3LO}(\mu_f,\mu_0,m_b,m_c) +\mathcal{O}(\alpha_s^4),
\end{align}
where $\mathcal{C}_s^{\rm N^3LO}(\mu_f,\mu,m_b,m_c)$ has dropped the $\mathcal{O}(\alpha_s^4)$ terms in Eq.~\eqref{eq:Cs01}. Namely the scale-dependence of the N$^3$LO matching coefficients rises
from $\mathcal{O}(\alpha_s^4)$, which is suppressed compared to lower-order.  However, the scale-independence coefficients of $\alpha_s^4$ such as $\mathcal{C}^{(3)}(x)$ and $\ln \mu_f$, which come from the $\mathcal{O}(\alpha_s^3)$ order in Eq.~\eqref{eq:Cs01}, are considerably large by aforementioned  calculation within the framework of  the NRQCD theory. These terms will lead to a significantly larger renormalization scale dependence at N$^3$LO. From Fig.~\ref{fig:fig3}, we also find the NRQCD factorization scale $\mu_f$ has a large influence on the matching coefficient.
When $\mu_f$ decreases, both the convergence of $\alpha_s$ expansion series  and the independence of $\mu$ will be improved.
The understanding of the observed large NNNLO corrections in this paper and the previous literatures~\cite{Egner:2022jot,Feng:2022ruy,Sang:2022tnh} is another important topic, which may
be related to the higher order corrections to NRQCD long-distance nonperturbative matrix elements, higher order relativistic corrections and the resummation techniques. We will leave it in the future work.

\section{Summary~\label{Summary}}

In this paper, we have performed  the $\rm N^3LO$ QCD corrections to the matching coefficient for the flavor-changing heavy quark scalar current  involving the bottom and charm quark
within the framework of the NRQCD effective theory.
For the first time, we obtain the analytical expressions of the three-loop renormalization constant and the corresponding three-loop anomalous dimension of the NRQCD scalar current
with different heavy quark mass $m_b$ and $m_c$.
Meanwhile, the three-loop matching coefficient $C_s(\mu_f,\mu,m_b,m_c)$ has also been obtained with high numerical accuracy, which  is helpful to analyze the threshold behaviours when two different heavy quarks
are close to each other.
The obtained $\rm N^3LO$ QCD  corrections are considerably large compared with lower-order corrections, and  exhibit
stronger dependence on the QCD renormalization scale and the NRQCD factorization scale at higher order, which suggests that   higher   order  QCD  corrections  should  be   combined   with  other techniques in order to get a  reliable prediction within the NRQCD  effective theory.

\begin{acknowledgments}

We would like to particularly thank J. H. Piclum for numerous helpful discussions. We also thank W.~L. Sang, X. Liu and X.~P. Wang for many useful discussions. This work is supported by NSFC under grant No.~11775117 and No.~12075124,  and by Natural Science Foundation of Jiangsu under Grant No.~BK20211267.

\end{acknowledgments}

\end{document}